UDC 004.056

# Investigating cybersecurity incidents using large language models in latest-generation wireless networks

**Leonid V. Legashev**

PhD, Leading Researcher, Laboratory of Digital Solutions and Big Data Analytics, Orenburg State University, Russia, 13 Prospect Pobedy, Orenburg, 460018. silentgir@gmail.com

**Arthur Yu. Zhigalov**

Junior Researcher, Laboratory of Artificial Intelligence and Data Analysis, Orenburg State University, Russia, 13 Prospect Pobedy, Orenburg, 460018 St. leroy137.artur@gmaill.com

**Abstract**

The purpose of research: Detection of cybersecurity incidents and analysis of decision support and assessment of the effectiveness of measures to counter information security threats based on modern generative models. The methods of research: Emulation of signal propagation data in MIMO systems, synthesis of adversarial examples, execution of adversarial attacks on machine learning models, fine tuning of large language models for detecting adversarial attacks, explainability of decisions on detecting cybersecurity incidents based on the prompts technique. Scientific novelty: A binary classification of data poisoning attacks was performed using large language models, and the possibility of using large language models for investigating cybersecurity incidents in the latest generation wireless networks was investigated. The result of research: Fine-tuning of large language models was performed on the prepared data of the emulated wireless network segment. Six large language models were compared for detecting adversarial attacks, and the capabilities of explaining decisions made by a large language model were investigated. The Gemma-7b model showed the best results according to the metrics Precision = 0.89, Recall = 0.89 and F1-Score = 0.89. Based on various explainability prompts, the Gemma-7b model notes inconsistencies in the compromised data under study, performs feature importance analysis and provides various recommendations for mitigating the consequences of adversarial attacks. Large language models integrated with binary classifiers of network threats have significant potential for practical application in the field of cybersecurity incident investigation, decision support and assessing the effectiveness of measures to counter information security threats.

**Acknowledgements**

This research is supported by RSF (grant 22-71-10124).

**Introduction.** As wireless networks proliferate and evolve, the amount of data transmitted over these networks increases dramatically. This increase in data volume enables the creation of new applications and services, but also leads to potential vulnerabilities that can be exploited by attackers. Common network threats include data poisoning attacks, wireless jamming attacks, attacks that intercept sensitive data, and attacks that masquerade as legitimate devices to gain unauthorized access to data. Modern artificial intelligence (AI) models have proven their effectiveness in detecting and mitigating such network threats [1]. It should be noted that AI models themselves are actively subjected to a large number of adversarial attacks, which reduce the performance of well-established models. The existing adversarial machine learning methodology allows for many different attacks such as poisoning, evasion, and many others, as noted in the study [2]. The field of artificial intelligence related to natural language processing (NLP) has received a new round of development in connection with the emergence of generative large language models (LLM), the main functionality of which is to generate a sequence of text tokens based on the input text to solve a wide range of applied problems [3].

**Literature review.** Many relevant studies are devoted to the issues of construction, compatibility and security in next-generation wireless networks. Baraboshin A. et al. [4] examine the problems and features of constructing systems for 6G wireless networks and compatibility with MIMO (massive Multiple Input Multiple Output) systems while ensuring high-speed data transmission. Hoang V. T. et al. [5] analyze the current state of research in the field of adversarial attacks on the latest generation wireless networks. Conclusions were made that attackers can successfully introduce noise interference into the signal, reducing the quality of classification when assessing the signal state. Shihab M. A. et al. [6] examine the use of adversarial learning, robust optimization and feature transformation to significantly increase the resilience of machine learning models to attempts to perform evasion type adversarial attacks. Khaleel Y. L. et al. [7] emphasize that in order to successfully counter the ever-growing network threats and new emerging types of network attacks, the means of protection and response to network threats must evolve at the same pace and operate in a proactive mode. Zhang W. et al. [8] study the classifiers of radio frequency signals and investigate their vulnerabilities to adversarial attacks, and also examine various practical issues related to the wireless environment, channel degradation, and mismatch between the attacker and transmitter signals. Son B. D. et al. [9] study the vulnerability of the 6G generation systems, including reconfigurable smart surfaces, massive MIMO systems, satellites, and semantic communications. Chiejina A. et al. [10] describe the developed prototype of the LTE/5G O-RAN (Open Radio Access Network) wireless testbed to evaluate the impact of adversarial attacks and the effectiveness of the distillation method as a protective measure. Within the framework of open radio access networks Ergu Y. A. et al. [11] perform a new type of adversarial attack that manipulates the parameters of the environment and degrades the user data transmission rate by up to 40 % and the packet delivery rate by up to 77.74 %. Zhang S. et al. [12] note that high-frequency components in wireless signals are a fundamental source of adversarial vulnerability and propose a homomorphic filtering algorithm that aims to effectively protect against adversarial samples by applying frequency domain filtering to the signal. Al Ghamdi M. A. [13] applies innovative graph structures and methods to protect wireless systems and mobile computing applications from adversarial inversion attacks that affect signal latency and throughput.

Recently, the application of large language models in the field of wireless networks has gained particular interest due to their outstanding capabilities in information understanding, analysis and problem solving. Zhang H. et al. [14] perform the task of automatic detection of DDoS attacks in a wireless network using large language models, and compare three contextual learning methods to improve the model performance. For the GPT-4 model, the accuracy and F1-Score metrics reach 90%. Shao J. et al. [15] present a comprehensive WirelessLLM framework for wireless communication systems, which includes operational design, advanced search generation, tool use, multimodal pre-training and fine-tuning depending on the task. Long S. et al. [16] optimize the network performance and intelligent network architecture based on a large language model, aim at creating a comprehensive intelligent 6G network system with the ability to capture patterns and highlight features in the transmitted data. Sheng Y. et al. [17] use large language models as a beam prediction method by formulating the millimeter wave (mmWave) beam prediction problem as a time series prediction problem where historical observations are aggregated and then converted into text representations using a trainable tokenizer. Xu S. et al. [18] study the development of universal fundamental models adapted to the unique needs of the latest generation wireless systems for the deployment of AI-based networks. A novel network intrusion prediction framework is proposed by Diaf A. et al. [19] that combines large language models for network traffic prediction with LSTM (Long short-term memory) networks for predicted traffic estimation. Chen Y. et al. [20] study various threat detection and monitor problems that can be addressed using large language models.

This paper will explore the potential of large language models to provide explainability and transparency to decisions made in detecting malicious network activity, which will contribute to the field of intelligent investigation of cybersecurity incidents in the latest generation of wireless networks.

**Perform an adversarial attack on the network data of an emulated wireless network segment.**

A basic approach for generating adversarial samples that can be used to attack machine learning models is the Fast Gradient Sign Method (FGSM) [21]. The idea behind this method is that it computes the gradients of the loss function with respect to the original data, and then uses the sign of the gradients to generate a new data sample that maximizes the loss *J* of the machine learning model:

$$x' = \varepsilon \cdot sign(\nabla_x J(\theta, x, y)),  \quad (1)$$

where $\varepsilon$ – minimum noise level, $\theta$ – the neural network model, $sign(\nabla_x J(\theta, x, y))$ – the gradient sign, $\nabla_x$ – the gradient, $x$ – the input data, $y$ – the target value for *x*.

DeepMIMO emulator [22] is used to generate accurate ray tracing data depending on the geometry, materials of the environment, and the locations of transmitters and receivers. In our previous study [23], we generated the "Boston5G_28" scenario data, which uses one base station fixed at a height of 15 meters and equipped with an omnidirectional antenna and two user array clusters with a total of 965,090 users located at a height of 2 meters. After running the "Boston5G_28" scenario, the following generated network activity features are available, which are presented in Table 1.

*Table 1*. Features extracted from the "Boston5G_28" scenario of the DeepMIMO emulator

| № | Feature | Description | Unit of measurement |
|---|---|---|---|
| 1 | X coordinate | X-coordinate of end user relative to emulated area | – |
| 2 | Y coordinate | Y-coordinate of end user relative to emulated area | – |
| 3 | Distance | Distance between base station and each user | meters |
| 4 | Pathloss | Combined path loss between sender and receiver (antenna signal attenuation) | decibels to 1 milliwatt |
| 5 | DoA_phi | Azimuth angle of signal arrival | degrees |
| 6 | DoA_theta | Zenith angle of signal arrival | degrees |
| 7 | DoD_phi | Azimuth angle of signal departure | degrees |
| 8 | DoD_theta | Zenith angle of signal departure | degrees |
| 9 | Phase | Phase of signal path | degrees |
| 10 | Power | Signal power at receiver | watts |
| 11 | Time of arrival | Signal arrival time | seconds |
| 12 | Line of Sight | Signal visibility status between base station and user, $LoS = \{-1, 0, 1\}$ | – |

The target feature of prediction is the pathloss of combined signal losses. We perform an FGSM adversarial data poisoning attack on the regression model with the $\varepsilon = 1^{-10}$ and the fraction of attacked data *fract* = 0.99. Performing an adversarial attack with gradient sign maximization increases the value of the mean square error (MSE) metric by 33 % on average and reduces the value of the coefficient of determination $R^2$ by 10 %.

Figure 1 presents the results of the feature importance for predicting combined signal losses by a regression model based on the Shapley additive explanation model, which calculates the contribution of each feature to the model result. Features with positive SHAP values have a positive effect on the quality of the forecast, and features with negative SHAP values have a negative effect on the quality of the forecast. It should be noted that the location of end-user antennas outside the line of sight of the base station ($LoS \neq 1$) has a more negative effect on predicting combined signal losses compared to users in line of sight ($LoS = 1$). The signal arrival time values have the most positive effect on a good predicted result of combined signal losses, the opposite effect is observed for the signal power upon receipt.

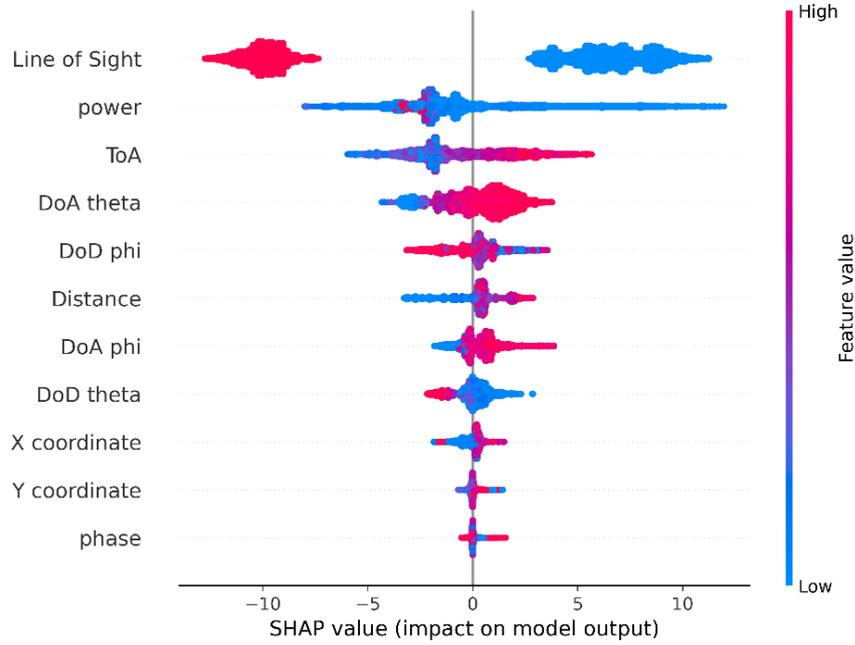

*Fig. 1*. SHAP model of the contribution of each feature to the model result

The resulting poisoned dataset will be used to solve a binary classification problem: determine if sample is regular data (Benign) with the label "0" or poisoned data (Malicious) with the label "1". Ensemble classifiers such as LightGBM solve the problem of binary classification of network threats with high accuracy, but the problem of the "blackbox" related to transparency and explainability of decisions made by the machine learning models remains relevant. To overcome the explainability barrier, we explore the capabilities of large language models in relation to the obtained emulated network data.

**Detecting adversarial attacks based on large language models.** Interaction with large language models usually consists of the following stages:

1. Fine-tuning of a large language model on prepared data to solve a specific problem.

2. Preparing instructions (prompt engineering) to obtain the correct output of the fine-tuned language model.

3. Evaluating the quality of the inference of the fine-tuned language model using quality metrics.

The fine-tuning process involves changing the initial parameters $\psi$ of the model to obtain new parameters $\tilde{\psi}$ that improve the inference of a large language model on a specific dataset:

$$\tilde{\psi} = \psi + \Delta\psi, \qquad (2)$$

where $\Delta\psi$ is an update of the model parameters. An update $\Delta\psi$ can be formalized as a function of the initial model parameters $\psi$, the dataset $D$ and the prompts $P$:

$$\Delta \psi = f(\psi, D, P), \qquad (3)$$

where $f$ – a fine-tuning algorithm that updates the model's parameters based on input data and prompts.

The data poisoned as a result of an adversarial attack and the original data in a 1:1 ratio is randomly divided into training (41837 records) and test (500 records) samples. The small number of test records is explained by the rather long output of responses from a large language model, in which token generation can take up to several seconds. For fine-tuning of a large language model, we will use the Unsloth library[1]. At the first stage, it is necessary to preprocess the data. Large language models work with text data in the form of a sequence of input tokens, so tabular data must be presented in a descriptive form. The training dataset $D$ should consist of three columns: ['**instruction**', '**input**', '**output**'], which corresponds to the roles of **system**, **user**, and **assistant**. In some cases, the '**instruction**' and '**input**' columns are combined into one column. The following instruction will be used as the prompt $P$ to a large language model: «*Some wireless network state records were compromised by an adversarial attack: values were changed so that the predicted signal pathloss value was incorrect. Based on the information provided about numerical features of a wireless signal, give answer if network traffic either (Benign) or (Malicious) and write your answer in round brackets*». As an input, a data row with features from Table 1 will be transformed into a descriptive text form according to a given template; one of two words will be an output: (Benign) or (Malicious). Figure 2 shows an example of such data row transformation.

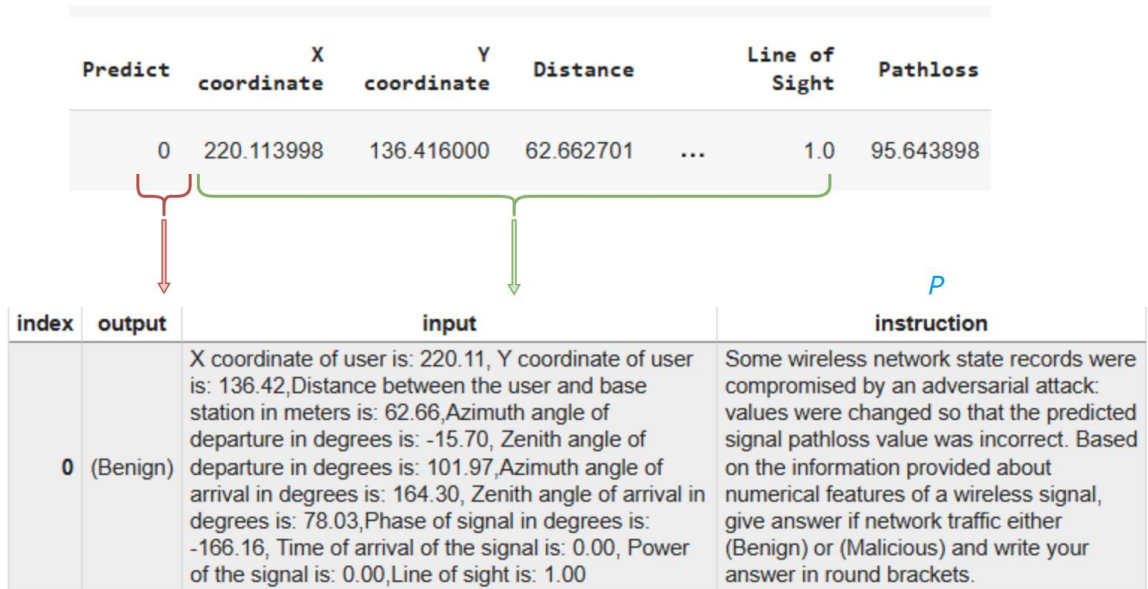

*Fig. 2*. Data transformation for fine-tuning of a large language model

To perform the fine-tuning process, we use the Supervised Fine-tuning Trainer as the function $f$. We set 200 training iterations, select AdamW-8bit as the optimizer, and set the maximum token sequence length *max_seq_length* = 2048 of the model outputs. We compare six lightweight large language models (up to eight billion parameters) by standard classification metrics Precision, Recall, and F1-Score, and the training loss. The results of binary classification of data poisoned by an adversarial attack are presented in Table 2.

---

[1] Unsloth – Daniel Han, Michael Han and Unsloth team, url= http://github.com/unslothai/unsloth, 2023

All models demonstrate acceptable accuracy, with the Gemma-7b model showing the best results on average in binary classification of compromised data.

*Table 2*. Comparison of quality metrics for poisoned data classification by fine-tuned large language models

| № | Model | Training loss | Precision | Recall | F1-score |
|---|---|---|---|---|---|
| 1 | unsloth/Meta-Llama-3.1-8B-bnb-4bit | 0,201 | 0,88 | 0,87 | 0,87 |
| 2 | mychen76/Llama-3.1_Intuitive-Thinker | 0,205 | 0,90 | 0,87 | 0,87 |
| 3 | unsloth/Qwen2.5-3B-Instruct | 0,199 | 0,86 | 0,86 | 0,86 |
| 4 | unsloth/gemma-7b | 0,182 | 0,89 | 0,89 | 0,89 |
| 5 | unsloth/mistral-7b-v0.3 | 0,162 | 0,90 | 0,87 | 0,87 |
| 6 | unsloth/DeepSeek-R1-Distill-Llama-8B-unsloth-bnb-4bit | 0,195 | 0,83 | 0,80 | 0,80 |

**Explainability of inference and investigation of cybersecurity incidents with large language models.** Large language models can overcome the "blackbox" problem of transparency of machine learning models' outputs to a large extent. We have prepared three variants of prompts that will be used as input for the fine-tuned large language model. In the instructions, we will ask the following questions:

– Based on the input data {input}, you proposed the answer {result}. Write your chain of thoughts and reasoning for this answer.

– Which of the input numerical features in {input} is the most important in identifying malicious activity?

– You are given a sample row of normal traffic data {input1} and a sample row of malicious traffic data {input2}. Analyze the sample rows and write your reasoning about how the adversarial attack affected the data.

Some responses from the Gemma-7b large language model are presented in Table 3. In its reasoning, the model notes that the distance between the user and the base station is significantly greater than the average value in the dataset, the azimuth and zenith angles of departure and arrival, and the phase difference between the signal and its arrival deviate significantly from the expected range for typical wireless signal propagation. The model notes that when performing adversarial attacks, changes in the data must be subtle enough to avoid detection, while significantly affecting the decision-making process of the AI system. The model typically highlights the distance between the user and the base station, the signal power, and the user's location in the line of sight of the base station as potentially significant features, which generally agrees with the results of the feature importance analysis of the SHAP model in Figure 1. When comparing two sample data records, the model provides recommendations for mitigating the consequences of adversarial attacks: implementing robust data verification and cleaning procedures; using anomaly detection algorithms to identify unusual patterns in the data; using diverse data sources to reduce the influence of biased data; training AI models using safe and reliable datasets.

*Table 3*. Inference of the Gemma-7b model for various explainability and reasoning validity prompts

| Gemma-7b model | |
|---|---|
| Prompt | *"Based on {input} info you predicted {result}. Please write your chain of thoughts and reasoning for such an answer."* |
| Answers | **(1)** *"\*\*Chain of Thoughts:\*\** |

|  | 1. **Statistical Analysis:** The distance between the user and base station is significantly higher than the average in the benign data set. |
|---|---|
|  | 2. **Domain Knowledge:** Such a drastic increase in distance is highly unusual and unlikely in normal wireless network conditions. |
|  | 3. **Abnormal Angles:** The azimuth and zenith angles of departure and arrival deviate significantly from the expected range for typical wireless signal propagation. |
|  | 4. **Phase Difference:** The phase difference between the signal and its arrival is outside the normal range, indicating potential manipulation. |
|  | 5. **Power Level:** The power level of the signal is zero, which suggests that the signal is artificially suppressed or altered. |
|  | **Conclusion:** |
|  | Based on these factors, the input data exhibits characteristics that deviate from the norm of legitimate wireless network data. Therefore, it is highly likely that the data is **Malicious**." |
|  | (2) "**Malicious activity** in this context can be defined as any action that intentionally alters or manipulates data presented to an AI system with the malicious intent of inducing inaccurate or harmful outputs. The modifications should be subtle enough to evade detection while significantly affecting the AI system's decision-making process. In the provided examples, the malicious data intentionally exaggerated the distance between the user and base station, interfered with signal strength, and obscured the line of sight, leading to potentially misleading interpretations by the AI system." |
| Prompt | "What is the most important numerical feature of {input} in determining malicious intent?" |
| Answers | (1) "Determining the most important numerical feature for identifying malicious intent requires further analysis and knowledge about the specific AI system and its training data. However, some features mentioned in the text might be potentially significant: |
|  | **Distance between the user and base station:** This feature significantly increased in the malicious data set, suggesting potential manipulation. |
|  | **Power of the signal:** The malicious data set assigned a power level of -0.00, which is unusual and could indicate interference. |
|  | **Line of sight:** The malicious data set shows "Line of sight" as -0.00, indicating potential obstruction between the user and base station. |
|  | These features might be highly relevant depending on the AI system's training and the specific attack techniques employed. Further investigation and domain knowledge are needed to determine the most impactful numerical features in this context." |
|  | (2) "**The adversarial attack significantly affected the numerical features of the input examples:** |
|  | ** Distance between the user and base station: The malicious data showed a significantly smaller distance, suggesting spoofing. |
|  | ** Azimuth and zenith angles: Notable discrepancies in these angles in the malicious data indicate signal manipulation. |
|  | ** Phase of the signal: A significant phase difference between the signals in the malicious data suggests signal manipulation. |
| | *** Power of the signal: The malicious data showed suppressed signal power, indicating data suppression."* |
| Prompt | *"Here is an example of a row of (Benign) traffic data {input1} and here is an example of a row of (Malicious) traffic data {input2}. Analyze both examples and write your thoughts on how adversarial attack affected data."* |
| Answers | **(1)** *"## Analysis of Adversarial Attack on Traffic Data*

***Comparison of Benign and Malicious Data:***

* ***Geographical location:*** *The malicious data shifted the user's location significantly further away from the base station compared to the benign data.*

* ***Distance to base station:*** *The malicious data increased the distance between the user and the base station by more than double.*

* ***Signal strength:*** *The malicious data set assigned a power level of -0.00, while the benign data was 0.00.*

* ***Line of sight:*** *The malicious data set shows "Line of sight" as -0.00, indicating potential obstruction between the user and base station.*

*** Possible Effects of Adversarial Attack:***

* ***Manipulation of Location:*** *The attacker could potentially manipulate the user's location data to mislead the AI system about the user's actual position.*

* ***Fabrication of Distance:*** *The increased distance between the user and base station could trigger different routing or authentication protocols.*

* ***Signalling Interference:*** *The malicious data might resemble noise or interference, leading the AI system to inaccurate interpretations of the signal.*

* ***Compromised Security:*** *The AI system might misclassify legitimate users due to the adversarial modifications.*

*** Potential Consequences of Adversarial Attacks:***

* ***Degraded Performance:*** *The AI system could produce inaccurate results, leading to poor network optimization or security vulnerabilities.*

* ***Manipulation of Services:*** *Attackers could potentially manipulate user experiences or access to services based on biased data interpretations.*

* ***Financial Losses:*** *Financial systems relying on AI for decision-making could suffer losses due to malicious data manipulation.*

*** Recommendations to Mitigate Adversarial Attacks:***

* *Implement robust data validation and cleaning procedures.*

* *Employ anomaly detection algorithms to identify unusual data patterns.*

* *Use diverse data sources to reduce the impact of biased data.*

* *Train AI models with secure and reliable data sets."*

**(2)** *"## Analysis of Malicious Data:*

***1. Distance:***

*- The distance between the user and base station is drastically higher than the average in benign data.*

*- This suggests an artificially inflated signal distance.* |

| | *\*\*2. Angles:\*\** |
| --- | --- |
| | *- Azimuth and zenith angles deviate significantly from normal ranges, indicating manipulation of signal propagation path.* |
| | *- Such anomalies could be achieved through signal reflections or interference.* |
| | *\*\*3. Phase Difference:\*\** |
| | *- The phase difference between the signal and its arrival lies outside the expected range, hinting at deliberate alteration of the signal.* |
| | *\*\*4. Power Level:\*\** |
| | *- The power level of the signal is zero, suggesting suppression or manipulation of the original signal.* |
| | *## Impact of Adversarial Attack:* |
| | *The adversarial attack in this case involved:* |
| | *\* \*\*Signal Spoofing:\*\* Distorting the signal's origin and trajectory, leading to inaccurate distance measurements.* |
| | *\* \*\*Reflection Manipulation:\*\* Deliberately reflecting the signal off surfaces, altering the angles of arrival and departure.* |
| | *\* \*\*Signal Suppression:\*\* Suppressing the signal's amplitude to zero, making it undetectable.* |
| | *These techniques manipulate the fundamental characteristics of the wireless signal, leading to the observed anomalies in the data.* |
| | *## Conclusion:* |
| | *The observed data exhibits significant deviations from benign network traffic patterns, indicating the presence of an adversarial attack. The specific techniques employed involved signal spoofing,* |
| | *reflection manipulation, and signal suppression."* |

**Conclusion.** In this study we compared six large language models of binary classification of adversarial data poisoning attacks. The best results were shown by the Gemma-7b model with the achieved metric values of Precision = 0.89, Recall = 0.89 and F1-Score = 0.89. In terms of explainability and reasoning, the Gemma-7b model notes inconsistencies in the studied data, analyzes the features importance and provides recommendations for mitigating the consequences of adversarial attacks. Large language models in the area of classifying malicious network activity are inferior to ensemble classifiers in terms of accuracy and resource intensity, but allow partially overcoming the barrier of explainability and transparency of the output of artificial intelligence models. The integration of fine-tuned language models together with binary classifiers directly in the network infrastructure will allow for a faster and more accurate response to emerging network anomalies, recreate explanations for each decision made by the target model, as well as investigate decision support and evaluate the effectiveness of measures to counter information security threats.

**References**

1. Zuev V.N. Network anomalies detection by deep learning // Software & Systems – 2021. – №1 (34) – P. 91–97. DOI: 10.15827/0236-235X.133.091-097

2. Korneev N.V., Kotrini E.S. Pattern for Ensuring Application Security under the Threat of Modifying a Machine Learning Model // Cybersecurity Issues. – 2025. – №. 1. – P. 117-127. DOI: 10.21681/2311-3456-2025-1-117-127

3. A. Golubinskiy, A.Tolstykh, M. Tolstykh. (2025). Automatic generation of scientific articles abstracts based on large language models. Informatics and Automation, 24(1), 275-301. DOI: https://doi.org/10.15622/ia.24.1.10

4. Baraboshin A. Yu., Luchin D. V., Maslov E. N. Analysis of application requirements and technological capabilities of radio signals promising for 6G networks // Cybersecurity Issues. – 2024. – №. 4. – P. 45-56. DOI: 10.21681/2311-3456-2024-4-45-56

5. Hoang V. T. et al. Security risks and countermeasures of adversarial attacks on AI-driven applications in 6G networks: A survey // Journal of Network and Computer Applications. – 2024. – P. 1-31. DOI: https://doi.org/10.1016/j.jnca.2024.104031

6. Shihab M. A. et al. Towards Resilient Machine Learning Models: Addressing Adversarial Attacks in Wireless Sensor Network // Journal of Robotics and Control (JRC). – 2024. – V. 5. – №. 5. – P. 1599-1617. DOI: https://doi.org/10.18196/jrc.v5i5.23214

7. Khaleel Y. L. et al. Network and cybersecurity applications of defense in adversarial attacks: A state-of-the-art using machine learning and deep learning methods // Journal of Intelligent Systems. – 2024. – V. 33. – №. 1. – P. 1-45. DOI: https://doi.org/10.1515/jisys-2024-0153

8. Zhang W., Krunz M., Ditzler G. Stealthy adversarial attacks on machine learning-based classifiers of wireless signals // IEEE Transactions on Machine Learning in Communications and Networking. – 2024. – V. 2. – P. 261-279. DOI: 10.1109/TMLCN.2024.3366161

9. Son B. D. et al. Adversarial attacks and defenses in 6g network-assisted IoT systems // IEEE Internet of Things Journal. – 2024. DOI: 10.1109/JIOT.2024.3373808

10. Chiejina A. et al. System-level analysis of adversarial attacks and defenses on intelligence in O-RAN based cellular networks // Proceedings of the 17th ACM Conference on Security and Privacy in Wireless and Mobile Networks. – 2024. – P. 237-247. DOI: https://doi.org/10.1145/3643833.365611

11. Ergu Y. A. et al. Unmasking Vulnerabilities: Adversarial Attacks Against DRL-based Resource Allocation in O-RAN // ICC 2024-IEEE International Conference on Communications. – IEEE, 2024. – P. 2378-2383. DOI: 10.1109/ICC51166.2024.10623131

12. Zhang S. et al. HFAD: Homomorphic filtering adversarial defense against adversarial attacks in automatic modulation classification // IEEE Transactions on Cognitive Communications and Networking. – 2024. – V. 10. – №. 3. – P. 880-892. DOI: 10.1109/TCCN.2024.3360514

13. Al Ghamdi M. A. Analyze textual data: deep neural network for adversarial inversion attack in wireless networks // SN Applied Sciences. – 2023. – V. 5. – №. 12. – P. 386. DOI: https://doi.org/10.1007/s42452-023-05565-8

14. Zhang H. et al. Large language models in wireless application design: In-context learning-enhanced automatic network intrusion detection // arXiv preprint arXiv:2405.11002. – 2024. DOI: https://doi.org/10.48550/arXiv.2405.11002

15. Shao J. et al. WirelessLLM: Empowering large language models towards wireless intelligence // arXiv preprint arXiv:2405.17053. – 2024. DOI: https://doi.org/10.48550/arXiv.2405.17053

16. Long S. et al. 6G comprehensive intelligence: network operations and optimization based on Large Language Models // IEEE Network. – 2024. DOI: 10.1109/MNET.2024.3470774

17. Sheng Y. et al. Beam prediction based on large language models // IEEE Wireless Communications Letters. – 2025. DOI: 10.1109/LWC.2025.3543567

18. Xu S. et al. Large multi-modal models (LMMs) as universal foundation models for AI-native wireless systems // IEEE Network. – 2024. DOI: 10.1109/MNET.2024.3427313

19. Diaf A. et al. Beyond detection: Leveraging large language models for cyber attack prediction in IoT networks // 2024 20th International Conference on Distributed Computing in Smart Systems and the Internet of Things (DCOSS-IoT). – IEEE, 2024. – P. 117-123. DOI: 10.1109/DCOSS-IoT61029.2024.00026

20. Chen Y. et al. A survey of large language models for cyber threat detection // Computers & Security. – 2024. – P. 104016. DOI: https://doi.org/10.1016/j.cose.2024.104016

21. Liu Y. et al. Sensitivity of adversarial perturbation in fast gradient sign method // 2019 IEEE symposium series on computational intelligence (SSCI). – IEEE, 2019. – P. 433-436. DOI: 10.1109/SSCI44817.2019.9002856

22. Alkhateeb A. DeepMIMO: A generic deep learning dataset for millimeter wave and massive MIMO applications // arXiv preprint arXiv:1902.06435. – 2019. DOI: https://doi.org/10.48550/arXiv.1902.06435

23. Legashev L. V., Zhigalov A. Yu. Study of Adversarial Attacks on Regression Machine Learning Models in 5G Wireless Networks // Cybersecurity Issues. – 2024. – №. 3. – P. 61-67. DOI: 10.21681/2311-3456-2024-3-61-67